\documentclass{edp-conf}
\usepackage{graphicx}
\begin{document}

\TitreGlobal{SF2A 2001}

%%-----------------------------
%%      the top matter
%%-----------------------------
\title{Secular evolution in the central regions of galaxies} 
\author{Eric Emsellem}\address{Centre de Recherche Astronomique de Lyon, 9 av. Charles
Andr\'e, 69561 Saint-Genis Laval, France}
\maketitle
\begin{abstract} 
In this paper, I mention a few processes which may play a role
in the evolution of the central regions of galaxies. In this context,
I briefly discuss some issues regarding the formation of bulges in spirals,
the role of supermassive black holes, and the importance of 
nuclear density waves.
\end{abstract}
%
%%-----------------------------
%%      your text
%%-----------------------------
\section{Introduction}

This paper is not meant as a thorough survey of our present
knowledge of the evolution processes involved in the central
regions of galaxies, but more as a short commentary on some
of the interesting issues which were recently discussed in
the literature. What are the actual physical processes that do 
act on the central structures and dynamics of galaxies? I will focus here on the search 
for the traces that these processes leave and which can be studied in nearby ''normal'' galaxies.
I will thus start with a short discussion on bulge properties
and formation scenarios, and then shortly report on the present situation
regarding central cusps and the role of supermassive black holes.
I will then briefly mention the importance of nuclear density waves,
spirals, bars and $m=1$ modes, in the evolution of the central
structures.

\section{Spiral Galaxies}

\begin{figure}[h]
\centering
\resizebox{6.2cm}{!}{\includegraphics[clip]{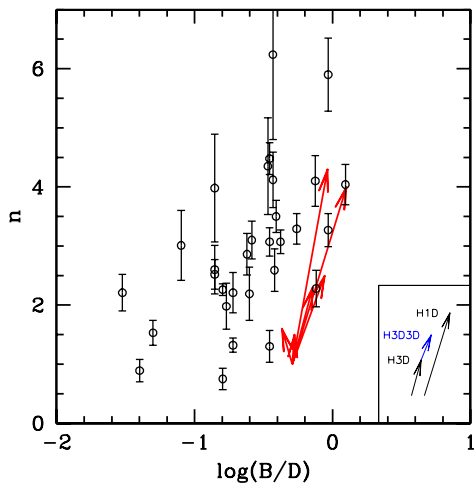}}
\caption{Growth vectors in the [$n$ - log(bulge to disk ratio$\equiv B/D$)] plane due
to the accretion of a small satellite galaxy. Each arrow
starts at the location of the original model and ends at the $n$ and $B/D$ derived
from a fit to the surface density profile after the merger. Point with error bars
are actual observations reported in Andredakis \etal\ (1995). See Aguerri \etal\ for
more details (their Figure~6).}
\label{fig:aguerri}
\end{figure}

When studying the central regions of spiral galaxies,
it is usually difficult to disentangle the relative
contributions of the bulge and disk. The bulge component
is generally assumed to dominate the central light profiles,
often parametrized as a Sersic law $\mu(r) \propto r^{1/n}$
(Andredakis, Peletier \& Balcells 1995; Fig~.\ref{fig:aguerri}): there is a tendency for
$n$ to increase from about 1 (exponential law) for late type
spirals to values around 4-6 (closer to a de Vaucouleurs profile) for early-type spirals.
Recent studies (e.g. Peletier \etal\ 1999,
Carollo \etal\ 2001), benefiting from the high
spatial resolution of HST, show that early-type bulges tend
to be older, with a small spread in age, similarly to ellipticals. 
Suggestions followed that exponential bulges in late-type galaxies,
having scaling properties (e.g. $\mu_e$/R$_e$) typical of disks,
may still be forming in the local universe through secular evolution of disks and bars
(Carollo 1999). Bulges growth via secular processes is often cited
as a way for spiral galaxies to evolve towards earlier Hubble type
(see Pfenniger, these Proceedings).
It was then argued that exponential bulges may not be able to evolve
into $r^{1/4}$-like bulges as they host central clusters massive enough
to prevent recurrent cycles of bar formation/disruption (Carollo 1999). 
However, the central regions could be dynamically cooled by later accretion of significant
amount of gas. Aguerri \etal\ (2001, see Fig.~\ref{fig:aguerri}) proposed that part
of the steepening of the surface brightness profiles could be due
to the accretion of dense satellites. This merging process 
heats the disk efficiently, which suggests that it did not play
a significant role in the making of exponential bulges in late-type galaxies.

%%%%%%%%%%%%%%%%%%%%%% FIGURE 1 - BEGIN
%%%%%%%%%%%%%%%%%%%%%% FIGURE 1 - END

\section{Cores of ellipticals and the role of the central mass
concentration}
Central surface brightness profiles of ellipticals, expressed
as $I \propto r^{-\gamma}$, seem to basically separate into two classes
(Faber \etal\ 1997): $\gamma < 0.3$ for luminous ellipticals, which 
have a clear change of slope in the inner regions, and are referred to as core galaxies;
and $\gamma > 0.5$ for low luminosity ellipticals, which have power-law cusps with no
clear breaks and are referred to as power-law galaxies. The transition between the two classes 
occurs around $M_B \sim -20.5$. Other properties, such as the internal dynamics (degree of isotropy)
or the isophote shape (disky/boxy), seem to correlate with the central value of $\gamma$.
Although the dichotomy between so-called core and power-law ellipticals
is clear, there are hints for the existence of an intermediate population of galaxies
which do not clearly belong to one of the two classes, and are
intermediate in terms of absolute magnitude (Rest \etal\ 2001). 
This tells us that our interpretation of the 
[density profile - $M_B$] plane for ellipticals may need to be refined
via a more adequate parametrization.
In any case, mechanisms leading to these stellar density profiles are not 
yet agreed upon: dissipative versus non-dissipative processes (Faber \etal\ 1997), 
adiabatic growth on a central black hole (van der Marel 1999),
diffusion due to binary black holes (Merritt \& Quinlan 1998)...

\begin{figure}[h]
\centering
\resizebox{6.2cm}{!}{\includegraphics[clip]{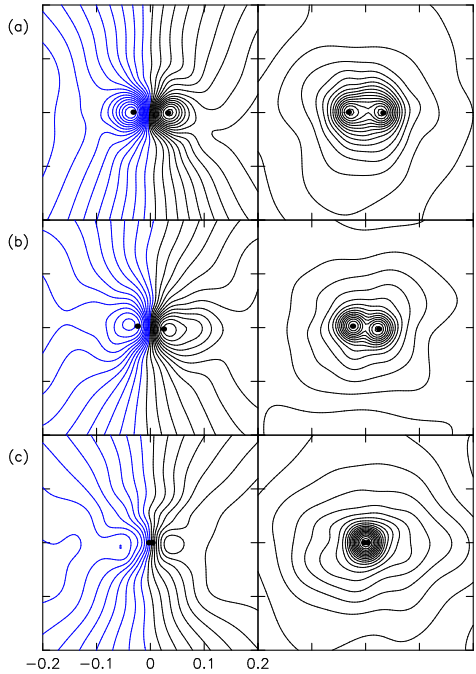}}
\caption{2D kinematics of the merging cusped galaxies for one of the model
of Milosavljevic \& Merritt (2001). View is in the plane of the merger from a direction
perpendicular to the line connecting the two black holes. Left panels: mean 
velocity. Right panels: velocity dispersion. See Milosavljevic \& Merritt (2001) for
more details (their Figure~6).}
\label{fig:milos}
\end{figure}

The ''binary black holes'' scenario has recently attracted more proponents, as 
new numerical studies of the merging of two galaxies, each containing
a central supermassive black hole, appeared in the literature
(see e.g. Milosavljevic \& Merritt 2001).
Although these simulations are still a long way from representing actual galaxy merging,
they show that a flattening of the central cusp slope can occur
with the hardening of the binary and ejection of stars.
Hope to observationally trace this process comes from 
the potential (but weak) dynamical signature left within the central
structure of the galaxy (e.g. attenuation of the circumnuclear rotation,
Fig.~\ref{fig:milos}). 
However, more work is required to implement some critical
issues into the game (gas, loss cone repopulation, mass spectrum, ...), 
hence to understand the true role of multiple black holes on the evolution of galactic nuclei.
%%%%%%%%%%%%%%%%%%%%%% FIGURE 2 - BEGIN
%%%%%%%%%%%%%%%%%%%%%% FIGURE 2 - END

Single central supermassive black holes, presumed to be present
in most galaxies, may also have their role in (re)shaping the central
dynamics, hence morphology, of the central regions. 
Chaotic diffusion may thus lead to a secular evolution of the orbital structure (Merritt 1999),
tending to axisymmetrise the galaxy from the centre, outwards.
Also, simulations performed by Holley-Bockelmann \& Richstone (2000)
showed that the presence of a central massive black hole may be critical 
to preserve the observed core fundamental relation for ellipticals 
(see above) during a non-equal mass merger.

\section{Density waves}
Density waves, which include $m=2$ modes such as spirals and bars,
and $m=1$ modes such as warps and lopsidedness, gained some audience as observations
showed they may be ubiquitous in the central regions of (disk) galaxies.

\subsection{$m=2$}
Bars have since long been recognised as potential actors for the redistribution
of angular momentum. But real proofs that this
was indeed happening in galaxies have been evasive.
Recently Sakamoto \etal\ (1999) have shown that molecular gas is more
concentrated in galaxies which are barred. The debate regarding the role of 
nuclear bars (secondary bar component within the large-scale primary bars) is still
open (Greusard \etal\ 2000; Erwin \etal\ 2001; Emsellem \etal\ 2001; Laine \etal\ 2001). Another class of actors, 
nuclear spirals, came up recently on the scene (Zaritsky \etal\ 1993; Regan \& Mulchaey 1999). 
At the moment of writing of this paper, it is too early
to decide whether these observed nuclear spirals have any direct or indirect
significance on the overall shaping of the central structures, but it
is an important avenue to consider.

\subsection{$m=1$}
Large-scale $m=1$ (lopsidedness) have long been ignored as possible evolution drivers.
But recent observational and theoretical studies showed that lopsidedness may
be a common mode in galaxies (Richter \& Sancisi 1994; Combes 2001 and references therein). 
These $m=1$ modes certainly do act on the redistribution
of the dissipative component, but, again, it is too early to conclude.

\subsubsection{Keplerian $m=1$}
\begin{figure}
\centering
\resizebox{6.2cm}{!}{\includegraphics[clip]{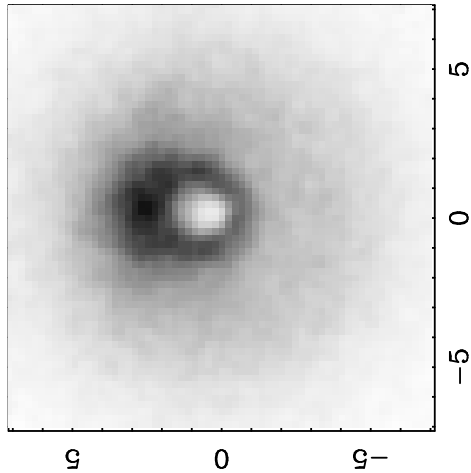}}
\caption{N-body simulation of an $m=1$ wave adapted to the case of M~31's nucleus.
View is face-on, scale is in parcsec. See Bacon \etal\ (2001) for more details.}
\label{fig:m31}
\end{figure}

At a very different scale, nuclear $m=1$ modes, which requires the gravitational potential to be 
close to keplerian, recently emerged in the context
of galactic nuclei (Fig.~\ref{fig:m31}), mainly triggered by the puzzling observations reported
on the nucleus of M~31 (see Bacon \etal\ 2001 and references therein). Although there is still some
uncertainty regarding the formation of such high amplitude waves (e.g. natural instability,
external perturbation), it seems that they are viable
modes, and may thus be of importance within the few central parsecs
where massive black holes are presumed to dominate the potential.
It is unfortunate that, at this scale, the stellar component is presently 
observable in only a handful of galaxies, but this situation can only improve.

%%%%%%%%%%%%%%%%%%%%%% FIGURE 3 - BEGIN
%%%%%%%%%%%%%%%%%%%%%% FIGURE 3 - END

\section{Conclusion}

There is no doubt that evolution occurs in the central regions
of galaxies, and this includes a wealth of complex processes,
a (very) few of which I have mentioned in this paper.
It is clear that density waves, in the form of nuclear bars, spirals,
or $m=1$, as well as massive black holes, do play a role
in shaping the central structures of galaxies. 
What also seems clear is that gas is a critical component in most of
the occuring mechanisms. 	

But at the moment, it is difficult, if not impossible, to have a
clear understanding of their relative contribution to the secular evolution
of central parts of galaxies. Timescales are short in the central kiloparsec,
and some of these processes may be recurrent (e.g. bar/spiral formation/destruction).
A step towards a more general understanding of these issues may need the implementation 
of a toy model, containing prescriptions for each of the 
thought-to-be important processes: see Combes (2001) for an attempt
along these lines. Such toy models, when fed into a typical galaxy merger tree,
may unravel some of the questions mentioned in this paper (although caution
is always the rule for such models).

%%-----------------------------
%%      your bibliography
%%-----------------------------

\end{document}